\documentclass[12pt]{article}
\usepackage{epsfig}
\begin{document}                   
%%%%%%%%%%%%%%%%%%%%%%%%%%%%%%%%%%%%%%%%%%%%%%%%%%%%%%%%%%%%%%%%%%%%%%%%%
\title{ANTIKAON ANGULAR DISTRIBUTIONS IN THE REACTION 
 ${\gamma}d \to K^-{\Theta}^+p \to K^-K^+np$ NEAR THE THRESHOLD AND
THE PARITY OF THE ${\Theta}^+$ PENTAQUARK}
\author{E.Ya. Paryev\\
{\it Institute for Nuclear Research, Russian Academy of Sciences,}\\
{\it Moscow 117312, Russia}}
%%==============================================================

\renewcommand{\today}{}

\maketitle

\begin{abstract}

        Within  spectator model we study the reaction
        ${\gamma}d \to K^-{\Theta}^+p \to K^-K^+np$ in the threshold
        energy region. We present the predictions for the exclusive and
        inclusive $K^-$-meson angular distributions in the laboratory 
        system for this reaction calculated for two possible parity states
        of the ${\Theta}^+$ resonance at 1.5 and $1.75~{\rm GeV}$ beam
        energies with and without imposing the relevant kinematical cuts
        on those parts of the sampled phase space where contribution from
        the main background sources, associated with the $\phi(1020)$,
        $\Lambda(1520)$ production as well as with the $K^-p$-rescattering
        in the final state, is expected to be dominant. We show that under
        chosen kinematics these distributions are sensitive to the
        ${\Theta}^+$ parity and, therefore, can be used as a filter for the
        determination of its parity.

\end{abstract}
                                                  
\newpage

\section*{1 Introduction}
   
   The study of an exotic pentaquark baryons
has received considerable interest in recent years
(see, for example, refs. [1--7], which contain a review of the experimental
and theoretical works on the issue) 
and is one of the most exciting topics of the
nuclear and hadronic physics nowadays. This interest was triggered by the
discovery of the narrow baryon resonance $\Theta^+(1540)$ with positive
strangeness $S=+1$ by the LEPS Collaboration at SPring-8/Osaka [8] and the
subsequent other experiments [9--17]. The observed state $\Theta^+(1540)$
decays into a kaon and a nucleon and has been interpreted as $q^4{\bar q}$
pentaquark with quark structure $uudd{\bar s}$. Evidence for the existence
of another exotic pentaquark state $\Xi^{--}(1862)$ with mass $1.86~{\rm GeV}$,
width about $18~{\rm MeV}$ due to detector resolution, strangeness $S=-2$
and quark content $ddss{\bar u}$ has been reported by the NA49 Collaboration
at SPS [18]. In addition, the signal of a heavy pentaquark $\Theta_c(3099)$
in which the antistrange quark in the $\Theta^+$ is replaced by an anticharm
quark was found in recent experiment [19]. Meanwhile, there have been also
several experiments [20--27] at high energy in which no signals for those
pentaquark baryons have been observed. Moreover, no definite structure in
the $K^+n$ invariant mass spectrum from the reaction
${\gamma}p \to {\bar K}^{0}K^+n$ was observed at 1540 MeV in the recent
high--statistics and high--resolution experiment [28] indertaken by the
CLAS Collaboration at JLab. 
Therefore, the existence of these
baryons is still not completely established and more high--statistics
experiments with different beams, targets, energies are needed to obtain a
definite result for or against their existence.

 The mass of about $1.54~{\rm GeV}$ and decay width of less than
20--$25~{\rm MeV}$ of the $\Theta^+$, extracted from the experiments [8--17],
are compatible with theoretical predictions of the chiral soliton model [29].
The observed $\Theta^+$ width reflects the experimental resolutions and its
actual magnitude, as is expected [30--35] from the analysis of the 
kaon--nucleon, kaon--deuteron and kaon--nucleus scattering data, is limited
by a few MeV level. While the isospin of the $\Theta^+$ resonance is probably
zero (see, e.g., SAPHIR [11] and CLAS [12] results concerning non--existence
of the $\Theta(1540)$ in $K^+p$ channel), the other quantum numbers of this
state including spin and parity have not yet been determined experimentally.
Theoretically, most models predict that $\Theta^+$ has spin $1/2$ because of
its low mass, whereas their predictions on the $\Theta^+$ parity are still
controversial. Thus, for example, the positive parity of the $\Theta^+$ is
supported by the chiral soliton model [29, 36, 37], various correlated quark
models [38--42], Skyrme model [43], and a lattice calculation [44]. On the
other hand, such theoretical approaches as the uncorrelated quark model [45],
the collective stringlike model of pentaquarks [46], 
the QCD sum rules [47--49], the lattice QCD [50, 51] favor a negative parity
for the $\Theta^+(1540)$. So, it is unclear currently what sign of the 
$\Theta^+$ parity is the correct one. The knowledge of this sign is important
for distinguishing between different models mentioned above and, hence, for
gaining more insight into the dynamics of low--energy QCD [52].

    To help determine the parity of the $\Theta^+$, a number of studies have
been carried out to understand how the unpolarized [53--66] and polarized
[59, 62--77] observables of the $\Theta^+$ production processes, induced by
the medium energy photons, nucleons, pions, and kaons on nucleon targets,
depend on the parity of $\Theta^+(1540)$. Very recently, the authors of [78]
have explored how the spin observables in the reaction
$\pi^{\pm}{\vec D} \to {\vec \Sigma}^{\pm}\Theta^+$ near the threshold can be
used to distinguish the parity of $\Theta^+$. Obviously, the use of the 
unpolarized observables for the determination of the $\Theta^+$ parity, which
do not depend very much on the theoretical ambiguities (if such observables
exist), has an advantage compared to the utilizing of the spin ones, since
in the first case a much simpler experimental setups and beam conditions are
required for the measurements. Recently, in refs. [58, 59], the authors 
discussed a rather model--independent way to discriminate the $\Theta^+$
parity from the ${\gamma}N \to {\bar K}\Theta^+$ reaction by looking at
antikaon angular distribution. In particular, they have demonstrated that the
(unpolarized) differential cross section for the reaction 
${\gamma}n \to K^-\Theta^+$ close to the production threshold shows a clear
distinction between the two opposite parities of the $\Theta^+$ baryon.
Namely, near the threshold
\footnote
{At excess energies above the $K^-\Theta^+$ threshold less than approximately
100 MeV as may be inferred from the arguments presented in the work [59], or,
respectively, at the photon energies smaller than about 2 GeV if the reaction
${\gamma}n \to K^-\Theta^+$ takes place on a free target neutron being at rest.}
, this cross section is isotropic in the ${\gamma}n$ c.m.s. frame if the parity
of the $\Theta^+$ is positive, and it follows 
$\sin^2{\theta_{K^-}^{'}}$ behavior
(where $\theta_{K^-}^{'}$ is 
the $K^-$-meson polar production angle in the c.m.s.)
when the parity of $\Theta^+$ is negative. Therefore, measurement of the
reaction ${\gamma}n \to K^-\Theta^+$ in the threshold energy region would allow
one to determine the parity of the $\Theta^+$ resonance [58, 59]. However,
such measurement can be performed only on the bound in the nucleus neutron 
because of the absence of a free neutron target. Often, the bound neutron in
the deuteron is used as a substitute of the free one. Thus, for instance, at
JLab, the $\Theta^+$ baryon was observed with the CLAS detector [10] as a
narrow peak in the $K^+n$ system produced in the reaction
${\gamma}n \to K^-\Theta^+ \to K^-K^+n$, where the target neutron was bound
in the deuteron. Unfortunately, the neutron in the deuteron is not at rest and
is moving with a Fermi momentum which has a component along the incident photon
direction of say about ${\pm}$ 50 MeV/c. Though this is only a few MeV in
energy, it has a huge influence on the kinematics, especially, if we are 
investigating the threshold phenomena. This raises the question of whether a
predicted [58, 59] specific shape of the angular distribution of the 
${\gamma}n \to K^-\Theta^+$ reaction near threshold, depending on the 
$\Theta^+$ parity, survives when this reaction takes place on the moving 
neutron in the deuteron. It is highly desirable and useful to give a reasonable
answer to this question--the main goal of the present investigation--to clarify
the feasibility of experimental determination of the parity of an exotic
pentaquark baryon state $\Theta^+$ by measuring this distribution. In doing so,
it is needed also to take into consideration the fact that the two--body 
reaction ${\gamma}n \to K^-\Theta^+$ is directly unobserved, since $\Theta^+$
can be detected only from its hadronic decays $\Theta^+ \to K^+n$ [8, 10--12]
and $\Theta^+ \to K^0p$ [9, 13--17].

     In this paper we perform a detailed analysis of the reaction
${\gamma}d \to K^-{\Theta}^+p \to K^-K^+np$ in the threshold energy region.
We present the predictions for the exclusive and inclusive antikaon angular
distributions in the laboratory system for this reaction obtained in the
framework of a simple spectator model for two possible parity states of the
$\Theta^+$ baryon at 1.5 and 1.75 GeV beam energies with and without imposing
the relevant kinematical cuts on those parts of the sampled phase space where
contribution from the main background sources, associated with the
$\phi(1020)$, $\Lambda(1520)$ production as well as with the
$K^-p$--rescattering in the final state, is expected to be dominant. We show
that under chosen kinematics these distributions are still sensitive to the
$\Theta^+$ parity and, therefore, can be used as an important tool for
identifying its parity.  

\section*{2 Spectator model} 

    Due to the high momentum transfer in the elementary process
${\gamma}n \to K^-\Theta^+$ near the threshold
\footnote
{Which amounts to $1.73~{\rm GeV}$ when the target neutron is at rest.}
and a large average separation of the neutron and proton in the deuteron
we can analyze the reaction ${\gamma}d \to K^-{\Theta}^+p \to K^-K^+np$
of our interest in the Impulse Approximation (IA) regime [79, 80]. In this
regime the reaction ${\gamma}d \to K^-{\Theta}^+p \to K^-K^+np$ reduces to 
the $\Theta^+$ photoproduction off the neutron in the deuteron:
%formula(1)
\begin{equation}
\gamma+n \to K^-+\Theta^+,
\end{equation}
and its subsequent decay into the $K^+n$
\footnote
{Because in the photoproduction experiments [8, 10--12] the $\Theta^+$ was
observed in the $K^+n$ decay mode, it is natural to consider this mode in the
present work.}
:
%formula(2)
\begin{equation}
\Theta^+ \to K^+n,
\end{equation}
while the recoiling proton acts as a spectator (see, fig. 1).
Considering that the width of $\Theta^+$ is very small compared to its mass
and using the results given in refs. [81, 82], we can represent in the IA the
differential cross section for creation of the four--body final state 
$K^-K^+np$ through the production/decay sequence (1, 2), taking place on a
neutron embedded in a deuteron, as follows:
%FORMULA (3)
\begin{equation}
d\sigma_{{\gamma}d \to K^-K^+np}^{(IA)}(E_{\gamma})=n_d(|{\bf p}_t|)
\delta({\bf p}_{t}+{\bf p}_{s})d{\bf p}_{t}d{\bf p}_{s}\times
\end{equation}
$$
\times
\frac{\pi}{I_2(s,m_{K},m_{\Theta^+})}
\frac{d\sigma_{{\gamma}n \to K^-{\Theta^+}}(s,{\theta}_{K^-}^{'})}
{d{\bf {\Omega}}_{K^-}^{'}}\times
$$
$$
\times
\delta({\bf p}_{\gamma}+{\bf p}_{t}-{\bf p}_{K^-}-{\bf p}_{\Theta^+})
\delta(E_{\gamma}+E_{t}-E_{K^-}-E_{\Theta^+})\frac{d{\bf p}_{K^-}}{E_{K^-}}
\frac{d{\bf p}_{\Theta^+}}{E_{\Theta^+}}\times
$$
$$
\times
\frac{d\Gamma_{\Theta^+ \to K^+n}(m_{\Theta^+},{\bf p}_{\Theta^+})}
{\Gamma_{\Theta^+}(m_{\Theta^+},{\bf p}_{\Theta^+})},
$$
where
%FORMULA (4)
\begin{equation}
I_2(s,m_K,m_{\Theta^+})=\frac{\pi}{\sqrt{s}}p_{K^-}^{'},
\end{equation}
%FORMULA (5)
\begin{equation}
p_{K^-}^{'}=\left|{\bf p}_{K^-}^{'}\right|=\frac{1}{2\sqrt{s}}
\lambda(s,m_{K}^{2},m_{\Theta^+}^{2}),
\end{equation}
%FORMULA (6)
\begin{equation}
\lambda(x,y,z)=\sqrt{{\left[x-({\sqrt{y}}+{\sqrt{z}})^2\right]}{\left[x-
({\sqrt{y}}-{\sqrt{z}})^2\right]}},
\end{equation}
%FORMULA (7)
\begin{equation}
s=\left(E_{\gamma}+E_{t}\right)^2-\left({\bf p}_{\gamma}+{\bf p}_{t}\right)^2,
\end{equation}
%FORMULA (8)
\begin{equation}
E_t=M_{d}-E_{s},\,\,\,E_s=\sqrt{{\bf p}_{s}^{2}+m_{p}^2},
\end{equation}
%FORMULA (9)
\begin{equation}
E_{K^-}=\sqrt{{\bf p}_{K^-}^{2}+m_{K}^2},\,\,
E_{\Theta^+}=\sqrt{{\bf p}_{\Theta^+}^{2}+m_{\Theta^+}^2};
\end{equation}
and
%FORMULA (10)
\begin{equation}
d\Gamma_{\Theta^+ \to K^+n}(m_{\Theta^+},{\bf p}_{\Theta^+})=
\frac{\left|M_{\Theta^+ \to K^+n}\right|^2}{2E_{\Theta^+}}(2{\pi})^4
\delta({\bf p}_{\Theta^+}-{\bf p}_{K^+}-{\bf p}_{n})\times
\end{equation}
$$
\times
\delta(E_{\Theta^+}-E_{K^+}-E_{n})\frac{d{\bf p}_{K^+}}{(2{\pi})^{3}2E_{K^+}}
\frac{d{\bf p}_{n}}{(2{\pi})^{3}2E_{n}},
$$
%FORMULA (11)
\begin{equation}
{\Gamma}_{\Theta^+}(m_{\Theta^+},{\bf p}_{\Theta^+})=
{\Gamma}_{\Theta^+}(m_{\Theta^+})/{\gamma}_{\Theta^+},\,\,
{\gamma}_{\Theta^+}=E_{\Theta^+}/m_{\Theta^+},
\end{equation}
%FORMULA (12)
\begin{equation}
E_{K^+}=\sqrt{{\bf p}_{K^+}^{2}+m_{K}^2},\,\,
E_{n}=\sqrt{{\bf p}_{n}^{2}+m_{n}^2}.
\end{equation}
Here, $(E_{\gamma},{\bf p}_{\gamma})$, $(E_{t},{\bf p}_{t})$,
$(E_{\Theta^+},{\bf p}_{\Theta^+})$, $(E_{K^-},{\bf p}_{K^-})$,
$(E_{K^+},{\bf p}_{K^+})$, $(E_{n},{\bf p}_{n})$, and
$(E_{s},{\bf p}_{s})$ are the four--momenta in the lab (or deuteron rest)
frame of the incoming photon, the struck target neutron, the intermediate
$\Theta^+$ resonance
\footnote
{Which is assumed to be on--shell, since its width is very small compared
to its mass.}
, the outgoing $K^-$, $K^+$-mesons and the neutron, and the recoil proton,
respectively; $d\sigma_{{\gamma}n \to K^-{\Theta^+}}(s,{\theta}_{K^-}^{'})/
d{\bf {\Omega}}_{K^-}^{'}$ is the off--shell
\footnote
{The struck target neutron is off--shell, see eq. (8).}
differential cross section for the production of a $K^-$-meson under the
polar angle ${\theta}_{K^-}^{'}$ with the momentum ${\bf p}_{K^-}^{'}$ in
reaction (1) in the ${\gamma}n$ c.m.s. 
(${\bf {\Omega}}_{K^-}^{'}={\bf p}_{K^-}^{'}/p_{K^-}^{'}$);
$n_d(|{\bf p}_t|)$ is the nucleon momentum distribution in the deuteron
normalized to unity; $m_p(m_n)$, $m_K$ and $M_d$ are the masses in free space
of a proton (neutron), kaon and deuteron, respectively; $m_{\Theta^+}$ is the
pole mass of the $\Theta^+$ baryon ($m_{\Theta^+}=1.54~{\rm GeV}$);
$\left|M_{\Theta^+ \to K^+n}\right|^2$ is the spin--averaged matrix element
squared describing the decay (2); ${\Gamma}_{\Theta^+}(m_{\Theta^+})$ is
the total width of the decay of $\Theta^+$ in its rest frame, taken at the
pole of the resonance.

      Let us now specify the off--shell differential cross section
$d\sigma_{{\gamma}n \to K^-{\Theta^+}}(s,{\theta}_{K^-}^{'})/
d{\bf {\Omega}}_{K^-}^{'}$ for $K^-$ production in the elementary process
(1), entering into eq. (3). Following refs. [79--82], we assume that this
cross section is equivalent to the respective on--shell cross section
calculated for the off--shell kinematics of the reaction (1). The on--shell
differential cross section for the reaction ${\gamma}n \to K^-{\Theta^+}$
has been calculated theoretically in refs. [58, 59] using both the
respective hadronic model and the CGLN amplitudes. The results of the
calculations show that this cross section in the threshold energy region,
i.e. at $E_{\gamma} \le 2~{\rm GeV}$, can be approximately parametrized by
%formula(13)
\begin{equation}
\frac{d\sigma_{{\gamma}n \to K^-{\Theta^+}}(s,{\theta}_{K^-}^{'})}
{d{\bf {\Omega}}_{K^-}^{'}}=\left\{
\begin{array}{ll}
	\frac{1}{4{\pi}}\sigma_{{\gamma}n \to K^-{\Theta^+}}^{(+)}(\sqrt{s})
	&\mbox{for the positive $\Theta^+$ parity}, \\
	&\\
        \frac{3}{8{\pi}}\sin^2{{\theta}_{K^-}^{'}}
        \sigma_{{\gamma}n \to K^-{\Theta^+}}^{(-)}(\sqrt{s})
	&\mbox{for the negative $\Theta^+$ parity}. 
\end{array}             
\right.	
\end{equation}
Here, $\sigma_{{\gamma}n \to K^-{\Theta^+}}^{(+)}(\sqrt{s})$ and
$\sigma_{{\gamma}n \to K^-{\Theta^+}}^{(-)}(\sqrt{s})$ are the on--shell 
total cross sections of the elementary process ${\gamma}n \to K^-{\Theta^+}$
for the positive and negative $\Theta^+$ parities, respectively. These cross
sections have been also calculated in refs. [58, 59] and the results of
calculations, taking into account that the $s$--wave ($p$--wave) antikaon
production
is expected [58, 59] near threshold when the $\Theta^+$ has the positive
(negative) parity, have been parametrized by us as follows:
%FORMULA (14)
\begin{equation}
\sigma_{{\gamma}n \to K^-{\Theta^+}}^{(+)}(\sqrt{s})=\frac{675p_{K^-}^{'}}
{1+2p_{K^-}^{'2}} [{\rm nb}],
\end{equation}
%FORMULA (15)
\begin{equation}
\sigma_{{\gamma}n \to K^-{\Theta^+}}^{(-)}(\sqrt{s})=\frac{595p_{K^-}^{'3}}
{1+15p_{K^-}^{'3}} [{\rm nb}],
\end{equation}
with $p_{K^-}^{'}$ denoting the $K^-$ three--momentum in the ${\gamma}n$
c.m.s. measured in GeV/c. This momentum is defined above by eq. (5).
An inspection of the formulas (14), (15) leads, as is easy to see, to the
conclusion that the total cross sections for the negative parity $\Theta^+$
are approximately 10--100 times smaller than those for the positive parity one
in the range of the photon energy 
$1.73~{\rm GeV} < E_{\gamma} < 3~{\rm GeV}$. Thus, for example, the positive
and negative $\Theta^+$ parity cases give the total cross sections of
$100~{\rm nb}$ and $2~{\rm nb}$, respectively, at $E_{\gamma}=1.8~{\rm GeV}$,
whereas at $E_{\gamma}=2.5~{\rm GeV}$ these cross sections, correspondingly,
are $230~{\rm nb}$ and $28~{\rm nb}$
\footnote
{It should be noted that these values are in disagreement with the 
results of the experiment [28]. In the light of these results the use
of eqs. (14), (15) enables us to obtain an upper estimate of the strength
of the respective antikaon angular distributions and has no influence on
their shape of our main interest.}
.

   The $K^-$-meson production angle ${\theta}_{K^-}^{'}$ in the ${\gamma}n$
c.m.s., entering into eq. (13), is defined by
%FORMULA (16)
\begin{equation}
\cos{{\theta}_{K^-}^{'}}=\frac{{\bf p}_{\gamma}^{'}{\bf p}_{K^-}^{'}}
{p_{\gamma}^{'}p_{K^-}^{'}},
\end{equation}
where ${\bf p}_{\gamma}^{'}$ denotes the three--momentum of an incident photon
in this system. Writting the relativistic invariant 
$t=[(E_{\gamma},{\bf p}_{\gamma})-(E_{K^-},{\bf p}_{K^-})]^2$ 
in the laboratory 
and in the ${\gamma}n$ c.m. systems and equating the results, we 
readily obtain
%FORMULA (17)
\begin{equation}
\cos{{\theta}_{K^-}^{'}}=\frac{p_{\gamma}p_{K^-}\cos{{\theta}_{K^-}}+
(E_{\gamma}^{'}E_{K^-}^{'}-E_{\gamma}E_{K^-})}
{p_{\gamma}^{'}p_{K^-}^{'}}.
\end{equation}
In the above, ${\theta}_{K^-}$ is the angle between the momenta
${\bf p}_{\gamma}$ and ${\bf p}_{K^-}$ in the lab frame, while
$E_{\gamma}^{'}$ and $E_{K^-}^{'}$ are the energies of the initial
photon and outgoing antikaon in the ${\gamma}n$ c.m.s., 
respectively.                
These energies are given by
%FORMULA (18)
\begin{equation}
E_{\gamma}^{'}=p_{\gamma}^{'}=\frac{1}{2\sqrt{s}}
\lambda(s,0,E_{t}^2-p_{t}^2),
\end{equation}
%FORMULA (19)
\begin{equation}
E_{K^-}^{'}=\sqrt{p_{K^-}^{'2}+m_{K}^2}.
\end{equation}

      Consider now the spin--averaged matrix element squared 
$\left|M_{\Theta^+ \to K^+n}\right|^2$ describing the decay 
$\Theta^+ \to K^+n$. Following the parity and angular momentum conservation
laws, the decay amplitude $M_{\Theta^+ \to K^+n}$ should exhibit a $p$-- or
$s$--wave behavior (for a spin--$\frac{1}{2}\Theta^+$) in the $\Theta^+$
rest frame when the $\Theta^+$ has the positive or negative parity,
respectively. However, if the spin state of the outgoing neutron is not fixed,
the difference between the angular distributions of the $\Theta^+ \to K^+n$
decay, corresponding to the positive and negative $\Theta^+$ parity,
disappears [65], which means that the spin--averaged matrix element squared
$\left|M_{\Theta^+ \to K^+n}\right|^2$ results in an isotropic angular
distribution of this decay for both parities of $\Theta^+$. By taking this
fact into consideration as well as integrating eq. (10) over the momenta
${\bf p}_{K^+}$ and ${\bf p}_{n}$ in the $\Theta^+$ rest frame, we can easily
get the following relation between $\left|M_{\Theta^+ \to K^+n}\right|^2$ and
the partial width $\Gamma_{\Theta^+ \to K^+n}(m_{\Theta^+})$ of the
$\Theta^+ \to K^+n$ decay: 
%FORMULA (20)
\begin{equation}
\frac{\left|M_{\Theta^+ \to K^+n}\right|^2}{(2{\pi})^2}=\frac{2m_{\Theta^+}^2}
{{\pi}\stackrel{*}p_{K^+}}\Gamma_{\Theta^+ \to K^+n}(m_{\Theta^+}),
\end{equation}
where
\footnote
{Note that the $K^+$ momentum $\stackrel{*}p_{K^+}$ in the $\Theta^+$ decay
into $K^+n$ in its rest frame, as is easy to calculate, is equal to
$269.7~{\rm {MeV/c}}$.}
%FORMULA (21)
\begin{equation}
\stackrel{*}p_{K^+}=\frac{1}{2m_{\Theta^+}}
\lambda(m_{\Theta^+}^2,m_{K}^2,m_{n}^2).
\end{equation}
By using the relation (20), one finds that the ratio \\
$d\Gamma_{\Theta^+ \to K^+n}(m_{\Theta^+},{\bf p}_{\Theta^+})/
\Gamma_{\Theta^+}(m_{\Theta^+},{\bf p}_{\Theta^+})$, entering into eq. (3), 
reduces to a simpler form:
%FORMULA (22)
\begin{equation}
\frac{d\Gamma_{\Theta^+ \to K^+n}(m_{\Theta^+},{\bf p}_{\Theta^+})}
{\Gamma_{\Theta^+}(m_{\Theta^+},{\bf p}_{\Theta^+})}=\frac{m_{\Theta^+}}
{{\pi}\stackrel{*}p_{K^+}}BR(\Theta^+ \to K^+n)
\delta({\bf p}_{\Theta^+}-{\bf p}_{K^+}-{\bf p}_{n})\times
\end{equation}
$$
\times
\delta(E_{\Theta^+}-E_{K^+}-E_{n})\frac{d{\bf p}_{K^+}}{2E_{K^+}}
\frac{d{\bf p}_{n}}{2E_{n}},
$$
where                    
%FORMULA (23)
\begin{equation}
BR(\Theta^+ \to K^+n)=
\Gamma_{\Theta^+ \to K^+n}(m_{\Theta^+})/\Gamma_{\Theta^+}(m_{\Theta^+}).
\end{equation}
According to [21, 83, 84], $BR(\Theta^+ \to K^+n)=1/2$ for both parities of
$\Theta^+$.

      Before going to the next step, we discuss now the nucleon momentum
distribution in the deuteron $n_d(p_t)$
\footnote
{Or the momentum distribution $n_d(p_s)$ of the nucleon--spectator produced
by the spectator mechanism in the reactions off the deuteron target nucleus.}
needed for our calculations. This momentum distribution has been calculated in
[85], using the Paris potential [86, 87], and the results of calculations
have been parametrized here by the simple analytical form (A1) (see also
formula (23) in ref. [82]). This form has been employed in our calculations of
the $K^-$ production cross sections in the reaction
${\gamma}d \to K^-{\Theta^+}p \to K^-K^+np$ reported in the paper. In fig. 2
we present the momentum distribution of the proton--spectator $p_s^2n_d(p_s)$
in this reaction (solid curve) calculated using the parametrization (A1) from
[85] for $n_d(p_s)$. It is clearly seen that this distribution has a sharp peak
with a maximum near $45~{\rm {MeV/c}}$ and the long tail above
$150~{\rm {MeV/c}}$. 

     Now, let us proceed to the identification of the kinematic regions where
the reaction ${\gamma}d \to K^-K^+np$, going via the production/decay
sequence (1, 2), is expected to dominate over the non--resonant background
\footnote
{The reactions which contribute to the same final state $K^-K^+np$ and do not
proceed through the virtual ${\Theta^+}$ state.}
.
It is natural to consider this reaction namely in these identified kinematic
regions. According to [8, 10, 65, 66], the main contribution to the
non--resonant background in the near--threshold region with 
$E_{\gamma} \le 2~{\rm GeV}$ comes from the intermediate $\phi$-meson and
$\Lambda(1520)$-hyperon photoproduction:${\gamma}N \to {\phi}N \to K^+K^-N$
and ${\gamma}p \to K^+{\Lambda(1520)} \to K^+K^-p$. Thus, for example, our
calculations
\footnote
{Performed in line with the formulas (25)--(31) given below.}
show that at $E_{\gamma}=1.8~{\rm GeV}$ the $K^+K^-$ invariant mass $M_{K^+K^-}$
in the process ${\gamma}n \to K^-{\Theta^+} \to K^-K^+n$ taking place on a free
neutron being at rest is distributed in the region
$1.0~{\rm GeV} \le M_{K^+K^-} \le 1.1~{\rm GeV}$. The narrow mass distribution
of the $\phi$ concentrates largely in the region of
the $K^+K^-$ invariant masses 
$1.00~{\rm GeV} < M_{K^+K^-} < 1.04~{\rm GeV}$ [8, 65] (the so--called "$\phi$
window") and, therefore, lies completely within the sampled kinematic region
indicated above, which makes the $\phi$-meson contribution to the respective
data sample significant [8, 10, 65, 66]. In order to suppress this contribution
and enhance signal to background ratio, the $\phi$-mesons have to be removed
from the data sample. In order to remove the $\phi$-mesons, events with
$1.00~{\rm GeV} < M_{K^+K^-} < 1.04~{\rm GeV}$ have to be rejected [8, 65].
This means that we have to eliminate the phase space with the $K^+K^-$
invariant mass from 1.00 to $1.04~{\rm GeV}$ in our consideration of the
reaction ${\gamma}d \to K^-{\Theta^+}p \to K^-K^+np$. To make this, we will
multiply the differential cross section (3) by the "$\phi$ phase space
eliminating" factor $Q(M_{K^+K^-})$ defined as: 
%formula(24)
\begin{equation}
Q(M_{K^+K^-})=\left\{
\begin{array}{ll}
	0
	&\mbox{for $1.00~{\rm GeV} < M_{K^+K^-} < 1.04~{\rm GeV}$}, \\
	&\\
        1
	&\mbox{otherwise}. 
\end{array}             
\right.	
\end{equation}
Before going further, one has to evaluate the invariant mass $M_{K^+K^-}$ of
a $K^+K^-$--pair produced in the production/decay sequence (1, 2). In order to
evaluate this quantity it is more convenient to put ourselves in the
${\gamma}n$ c.m.s. Then, the invariant $M_{K^+K^-}^2$ can be expressed through
the energies and momenta of the $K^+$ and the $K^-$, $E_{K^+}^{'}$,
${\bf p}_{K^+}^{'}$ and $E_{K^-}^{'}$, ${\bf p}_{K^-}^{'}$, in this system in
the following way:
%formula(25)
\begin{equation}
M_{K^+K^-}^2=\left(E_{K^+}^{'}+E_{K^-}^{'}\right)^2-
\left({\bf p}_{K^+}^{'}+{\bf p}_{K^-}^{'}\right)^2=2m_K^2+
2E_{K^+}^{'}E_{K^-}^{'}-2{\bf p}_{K^+}^{'}{\bf p}_{K^-}^{'},
\end{equation}
where
%formula(26)
\begin{equation}
E_{K^+}^{'}=\sqrt{{\bf p}_{K^+}^{'2}+m_{K}^2},
\end{equation}
and the quantities $p_{K^-}^{'}$ and $E_{K^-}^{'}$ are defined above by
eqs. (5) and (19), respectively. Taking into account that the kaon momentum
${\bf p}_{K^+}^{'}$ can be expressed via its momentum
${\bf {\stackrel{*}p}}_{K^+}$ in the $\Theta^+$ rest frame 
and the $\Theta^+$ momentum
${\bf p}_{\Theta^+}^{'}$ in the ${\gamma}n$ c.m.s. as [88]
%formula(27)
\begin{equation}
{\bf p}_{K^+}^{'}=\frac{p_{\Theta^+}^{'}\stackrel{*}E_{K^+}}{m_{\Theta^+}}
{\bf n}_{\Theta^+}+\stackrel{*}p_{K^+}\left\{{\bf n}_{K^+}^{*}+
({\gamma}_{\Theta^+}^{'}-1)\cos{{\theta}_{K^+}^{*}}{\bf n}_{\Theta^+}\right\},
\end{equation}
where
%formula(28)
\begin{equation}
\stackrel{*}E_{K^+}=\sqrt{\stackrel{*}p_{K^+}^{2}+m_{K}^2},
{\gamma}_{\Theta^+}^{'}=E_{\Theta^+}^{'}/m_{\Theta^+},
E_{\Theta^+}^{'}=\sqrt{{\bf p}_{\Theta^+}^{'2}+m_{\Theta^+}^2},
p_{\Theta^+}^{'}=p_{K^-}^{'},
\end{equation}
%formula(29)
\begin{equation}
{\bf n}_{\Theta^+}={\bf p}_{\Theta^+}^{'}/p_{\Theta^+}^{'},\,\,
{\bf n}_{K^+}^{*}={\bf \stackrel{*}p}_{K^+}/\stackrel{*}p_{K^+}, \,\,
\cos{{\theta}_{K^+}^{*}}={\bf n}_{K^+}^{*}{\bf n}_{\Theta^+},
\end{equation}
we easily get that:
%formula(30)
\begin{equation}
{\bf p}_{K^+}^{'2}=\left(\frac{p_{\Theta^+}^{'}\stackrel{*}E_{K^+}}
{m_{\Theta^+}}\right)^{2}+
\frac{2\stackrel{*}E_{K^+}E_{\Theta^+}^{'}\stackrel{*}p_{K^+}p_{\Theta^+}^{'}}
{m_{\Theta^+}^2}\cos{{\theta}_{K^+}^{*}}+
\end{equation}
$$
+
\stackrel{*}p_{K^+}^{2}
\left\{1+({\gamma}_{\Theta^+}^{'2}-1)\cos^2{{\theta}_{K^+}^{*}}\right\},     
$$
%formula(31)                          
\begin{equation}
2{\bf p}_{K^+}^{'}{\bf p}_{K^-}^{'}=-\frac{2p_{K^-}^{'}}{m_{\Theta^+}}
\left[p_{\Theta^+}^{'}\stackrel{*}E_{K^+}+
\stackrel{*}p_{K^+}E_{\Theta^+}^{'}\cos{{\theta}_{K^+}^{*}}\right].
\end{equation}
The $K^+$ momentum $\stackrel{*}p_{K^+}$ in the $\Theta^+$ rest frame, entering
into eqs. (27)--(31), is defined above by the (21). It should be emphasized
that, according to (5), (19), (25)--(31), the invariant mass $M_{K^+K^-}$
of interest depends only on the cosine of the $K^+$ decay angle
${\theta}_{K^+}^{*}$ in the $\Theta^+$ rest system and the squared invariant
energy $s$ available in the first--chance ${\gamma}n$--collision, which
simplifies the calculations presented below.

Similarly, our calculations
\footnote
{Carried out in line with the following absolute limits for the invariant
mass $M_{K^-p}$ of interest: 
$m_K+m_p \le M_{K^-p} \le \sqrt{(E_{\gamma}+M_d)^2-{\bf p}_{\gamma}^2}-
m_{\Theta^+}$.}
show that at $E_{\gamma}=1.8~{\rm GeV}$ the invariant mass $M_{K^-p}$ of the
$K^-p$ system in the reaction ${\gamma}d \to K^-{\Theta^+}p$ is distributed in
the region $1.432~{\rm GeV} \le M_{K^-p} \le 1.665~{\rm GeV}$ which straddles
the $\Lambda(1520)$ mass, since the peak corresponding to the $\Lambda(1520)$
lies basically [10] in the region of the $K^-p$ invariant masses
$1.485~{\rm GeV} < M_{K^-p} < 1.551~{\rm GeV}$. This  makes the $\Lambda(1520)$
contribution to the same final state of our interest significant [8, 10, 66].
To reduce this contribution and, respectively, to improve signal to background
ratio, the $\Lambda(1520)$ resonance has to be removed from the data sample by
rejecting events with [10] $1.485~{\rm GeV} < M_{K^-p} < 1.551~{\rm GeV}$.
This means that we have to eliminate also the phase space with the $K^-p$
invariant mass from 1.485 to 1.551 GeV in our study of the reaction
${\gamma}d \to K^-{\Theta^+}p \to K^-K^+np$.To do this, we will also multiply
the differential cross section (3) by the "$\Lambda(1520)$ phase space
eliminating" factor $Q(M_{K^-p})$. This factor is defined in the following way: 
%formula(32)
\begin{equation}
Q(M_{K^-p})=\left\{
\begin{array}{ll}
	0
	&\mbox{for $1.485~{\rm GeV} < M_{K^-p} < 1.551~{\rm GeV}$}, \\
	&\\
        1
	&\mbox{otherwise}. 
\end{array}
\right.	
\end{equation}
The invariant mass squared $M_{K^-p}^2$ can be obtained in a straightforward
manner, and the result is
%formula(33)
\begin{equation}
M_{K^-p}^2=\left(E_{K^-}+\sqrt{(-{\bf p}_t)^2+m_p^2}\right)^2-
\left({\bf p}_{K^-}-{\bf p}_{t}\right)^2=
\end{equation}
$$
=m_K^2+m_p^2+
2E_{K^-}\sqrt{(-{\bf p}_t)^2+m_p^2}+
2p_{K^-}p_{t}\cos{\theta_{{\bf p}_t{\bf p}_{K^-}}}
$$
with $\theta_{{\bf p}_t{\bf p}_{K^-}}$ being the angle between the momenta
${\bf p}_t$ and ${\bf p}_{K^-}$ in the lab frame. This angle is related
to the angles between ${\bf p}_{\gamma}$ and ${\bf p}_t$ ($\theta_t$),
${\bf p}_{\gamma}$ and ${\bf p}_{K^-}$ ($\theta_{K^-}$) and to the azimuthal
angles ${\varphi}_t$ of ${\bf p}_t$, ${\varphi}_{K^-}$ of ${\bf p}_{K^-}$ by
the trigonometric relation 
%FORMULA (34)
\begin{equation}
\cos{\theta_{{\bf p}_t{\bf p}_{K^-}}}=\cos{\theta_{K^-}}\cos{\theta_t}+
\sin{\theta_{K^-}}\sin{\theta_t}\cos{({\varphi}_t-{\varphi}_{K^-})}.
\end{equation}

   There is yet another background source, if we want to look at the antikaon
angular distributions from primary production process (1). This source of
background is related to the possible rescattering
\footnote
{Or final--state interaction (FSI).}
of the produced $K^-$-meson on the proton in the final state (see fig. 3).
Such rescattering may distort the angular distributions of antikaons produced
in ${\gamma}d$ interactions through the primary photon--induced reaction
channel ${\gamma}n \to K^-{\Theta^+}$ of interest (see fig. 1). Therefore,
we also need to specify the kinematic region, in addition to those specified
before, where the $K^-p$--rescattering is expected to be negligible. This
region has to be taken into consideration as well in the subsequent
calculations of the $K^-$ angular distributions from the primary reaction
channel (1). The effects of rescattering on the recoil nucleon in hadron--
and photon--deuteron interactions have been discussed previously (see, e.g.,
refs. [89--98] and references therein). Following [89, 90], the ratio of the
moduli of the amplitudes corresponding to the diagrams of fig. 3 and fig. 1
($M^{(FSI)}$ and $M^{(IA)}$, respectively) can be estimated using the relation
%FORMULA (35)
\begin{equation}                                                          
\frac{|M^{(FSI)}|}{|M^{(IA)}|} \approx \frac{|f_{K^-p \to K^-p}|}{4{\pi}R_d}
\frac{1}{q_{K^-p}R_d}\frac{{\varphi}_d(0)}{{\varphi}_d(p_s)}.
\end{equation}
Here, $f_{K^-p \to K^-p}$ is the elastic $K^-p$ scattering amplitude normalized
on the $K^-$ differential cross section  
$d\sigma_{K^-p \to K^-p}/d{\bf {\Omega}}_{c.m.s.}$ in the $K^-p$ c.m.s. by
$|f_{K^-p \to K^-p}|^2=d\sigma_{K^-p \to K^-p}/d{\bf {\Omega}}_{c.m.s.}$;
$q_{K^-p}$ is the relative momentum of the intermediate $K^-$-meson and the
spectator proton; $R_d$ is the average internucleon distance inside the
deuteron; ${\varphi}_d$ is the deuteron wave function in momentum space. The
$K^-p$ rescattering plays a significant role in the case of the relative
momenta $q_{K^-p}$ falling in the low--momentum region 
$q_{K^-p} < 100~{\rm {MeV/c}}$ where the $K^-p$ elastic cross section
$\sigma_{K^-p \to K^-p}$ is large 
($\sigma_{K^-p \to K^-p} > (80-100)~{\rm mb}$ [99--101]). Therefore, to reduce 
this rescattering we will restrict ourselves in the following to the case
of relative momenta $q_{K^-p} \ge 100~{\rm {MeV/c}}$
\footnote
{Accounting for the relation 
$q_{K^-p}=\frac{1}{2M_{K^-p}}\lambda(M_{K^-p}^2,m_K^2,m_{p}^2)$, we can easily
obtain that this corresponds to the region of the $K^-p$ invariant masses
$M_{K^-p} \ge 1.447~{\rm GeV}$.}
.
Then, employing, e.g.,  the Hulthen wave function
\footnote
{The quantity $R_d$ for this function is equal to $3.1~{\rm fm}$.}
[102] for ${\varphi}_d$ to estimate the ratio (35) and assuming that
$|f_{K^-p \to K^-p}| \approx \sqrt{80{\rm mb}/{4\pi}}$ here, one obtaines
that for $q_{K^-p} \ge 100~{\rm {MeV/c}}$ 
(or for $M_{K^-p} \ge 1.447~{\rm GeV}$) the contribution from the diagram
of fig. 3 is suppressed at least at the recoil proton momenta of 
$p_s < 280~{\rm {MeV/c}}$. 
Hence, the spectator mechanism of fig. 1 gives the dominant contribution to
the $\Theta^+$ photoproduction from the neutron in the deuteron at small
values of the spectator proton momentum $p_s$.

    So, the above considerations require that the $K^-p$ invariant mass must
be greater than $1.447~{\rm GeV}$ and the recoil proton momentum must be
smaller than $280~{\rm {MeV/c}}$. To fulfil these requirements, we will
multiply the differential cross section (3) by one more "$K^-p$ phase space
terminating" factor $Q(M_{K^-p},p_s)$. This factor is given by:
%FORMULA (36)
\begin{equation}
Q(M_{K^-p},p_s)={\theta}(M_{K^-p}-M_{cut}){\theta}(p_{cut}-p_s),
\end{equation}
where $M_{cut}=1.447~{\rm GeV}$, $p_{cut}=280~{\rm {MeV/c}}$ and $\theta(x)$
is the standard step function.

       Finally, by combining (3), (24), (32) and (36), we get within the IA
the following expression for the differential cross section of the reaction
${\gamma}d \to K^-{\Theta^+}p \to K^-K^+np$ over all the physical variables,
which includes the phase space cuts we introduced:
%formula(37) 
\begin{equation}
d\sigma_{{\gamma}d\to K^-K^+np}(E_{\gamma})=
d\sigma_{{\gamma}d\to K^-K^+np}^{(IA)}(E_{\gamma})Q(M_{K^+K^-})Q(M_{K^-p})
Q(M_{K^-p},p_s).
\end{equation}
Integrating eq. (37) over the available phase space with accounting for (22),
we get after some algebra the exclusive differential cross section of eq. (37)
in the laboratory frame of physical (and practical) interest, where the final
antikaon is detected without analyzing its energy for fixed three--momentum
of the proton--spectator:
%formula(38)
\begin{equation}
\frac{d\sigma_{{\gamma}d\to K^-K^+np}(E_{\gamma})}
{d{\bf {\Omega}}_{K^-}dp_sd{\bf {\Omega}}_s}=p_{s}^{2}n_d(p_s)
{\theta}(v_{K^-}^{'}-v_c){\theta}(p_{cut}-p_s)\times
\end{equation}
$$
\times
\frac{p_{K^-}^{(1)2}}{p_{K^-}^{'}
\sqrt{p_{K^-}^{'2}-{\gamma}_{c}^{2}v_{c}^{2}m_{K}^{2}
\sin^2{{\theta}_{K^-}^{c}}}}
\frac{d\sigma_{{\gamma}n\to K^-{\Theta}^+}
[s,{\theta}_{K^-}^{'}(p_{K^-}^{(1)})]}
{d{\bf {\Omega}}_{K^-}^{'}}\times
$$
$$
\times
Q[M_{K^-p}(p_{K^-}^{(1)})]
{\theta}[M_{K^-p}(p_{K^-}^{(1)})-M_{cut}]\times
$$
$$
\times
\frac{1}{2}BR({\Theta}^{+} \to K^+n)
\int\limits_{-1}^{1}Q[M_{K^+K^-}(p_t,{\theta}_t,\cos{{\theta}_{K^+}^{*}})]
d\cos{{\theta}_{K^+}^{*}}+
$$      
$$
+
p_{s}^{2}n_d(p_s){\theta}(v_c-v_{K^-}^{'}){\theta}(p_{cut}-p_s)\times
$$
$$
\times
\sum_{i=1}^2\frac{p_{K^-}^{(i)2}}{p_{K^-}^{'}
\sqrt{p_{K^-}^{'2}-{\gamma}_{c}^{2}v_{c}^{2}m_{K}^{2}
\sin^2{{\theta}_{K^-}^{c}}}}
\frac{d\sigma_{{\gamma}n\to K^-{\Theta}^+}
[s, {\theta}_{K^-}^{'}(p_{K^-}^{(i)})]}
{d{\bf {\Omega}}_{K^-}^{'}}\times
$$
$$
\times
Q[M_{K^-p}(p_{K^-}^{(i)})]
{\theta}[M_{K^-p}(p_{K^-}^{(i)})-M_{cut}]
\times
$$
$$
\times
\frac{1}{2}BR({\Theta}^{+} \to K^+n)
\int\limits_{-1}^{1}Q[M_{K^+K^-}(p_t,{\theta}_t,\cos{{\theta}_{K^+}^{*}})]
d\cos{{\theta}_{K^+}^{*}},
$$      
where
%FORMULA (39)
\begin{eqnarray}
{\bf {\Omega}}_{K^-}={\bf p}_{K^-}/p_{K^-},\,\,
{\bf {\Omega}}_{s}={\bf p}_{s}/p_{s},\,\, 
v_{K^-}^{'}=p_{K^-}^{'}/E_{K^-}^{'},\\ \nonumber
v_c=|{\bf v}_c|,\,\, 
{\bf v}_c=\frac{{\bf p}_{\gamma}+{\bf p}_t}{E_{\gamma}+E_t},\,\,
{\gamma}_c=\frac{1}{\sqrt{1-v_{c}^{2}}}; 
\end{eqnarray}
%FORMULA (40)
\begin{equation}
\cos{\theta_{K^-}^{c}}=\frac{{\bf p}_{K^-}{\bf v}_c}{p_{K^-}v_c}=
\frac{p_{\gamma}\cos{\theta_{K^-}}+p_t\cos{\theta_{{\bf p}_t{\bf p}_{K^-}}}}
{v_c(E_{\gamma}+E_t)},\,\, {\bf p}_t=-{\bf p}_s
\end{equation}
and
%FORMULA (41)
\begin{equation}
p_{K^-}^{(1)}=\frac{p_{K^-}^{'}(v_c/v_{K^-}^{'})\cos{{\theta}_{K^-}^{c}}+
\sqrt{p_{K^-}^{'2}-\gamma_{c}^{2}v_{c}^{2}m_{K}^{2}\sin^2{\theta_{K^-}^c}}}
{{\gamma}_c(1-v_{c}^2\cos^2{{\theta}_{K^-}^c})}\,\, for \,\, v_{K^-}^{'} > v_c,
\end{equation}
%FORMULA (42)
\begin{equation}
p_{K^-}^{(1, 2)}=\frac{p_{K^-}^{'}(v_c/v_{K^-}^{'})\cos{{\theta}_{K^-}^{c}}\pm
\sqrt{p_{K^-}^{'2}-\gamma_{c}^{2}v_{c}^{2}m_{K}^{2}\sin^2{\theta_{K^-}^c}}}
{{\gamma}_c(1-v_{c}^2\cos^2{{\theta}_{K^-}^c})}\,\, for \,\,
 v_{K^-}^{'} \le v_c.
\end{equation}                                                       
The quantity $\cos{\theta_{{\bf p}_t{\bf p}_{K^-}}}$, entering into eq. (40),
is defined above by the (34). It is worth noting that, as is evident from
eqs. (41), (42), in the case when the $K^-$-meson velocity $v_{K^-}^{'}$ in
the ${\gamma}n$ c.m.s. is greater than the velocity $v_c$ of this system in
the lab frame ($v_{K^-}^{'} > v_c$) the polar $K^-$ production angle
$\theta_{K^-}^{c}$ varies without restriction between 0 and $\pi$, otherwise
($v_{K^-}^{'} \le v_c$) it lies in the interval 
$(0, \arcsin{(p_{K^-}^{'}/{\gamma_c}{v_c}m_K)})$. Thus, for instance, simple
calculations show that the maximal value of the $K^-$-meson production angle
$\theta_{K^-}$ in the lab system in the reaction ${\gamma}n \to K^-\Theta^+$
taking place on a free neutron being at rest
\footnote
{In this case, as is easy to see, $\theta_{K^-}=\theta_{K^-}^{c}$.}  
at $E_{\gamma}=1.8$ GeV amounts approximately to 21$^{0}$, i.e., in the
threshold energy region antikaons are mainly emitted in this reaction in
forward directions (see, also, figs. 4--8 given below). Further observables
of our interest for the exclusive $d(\gamma, K^-p)K^+n$ and inclusive
$d(\gamma, K^-)K^+np$ processes, proceeding through the intermediate $\Theta^+$
state, are the $K^-$-meson angular distributions, respectively, for fixed and
non--fixed solid angle of the proton--spectator. Because of eq. (38), they are
given by:
%formula(43) 
\begin{equation}
\frac{d\sigma_{{\gamma}d\to K^-K^+np}(E_{\gamma})}
{d{\bf {\Omega}}_{K^-}d{\bf {\Omega}}_{s}}=
\int{dp_{s}}\frac{d\sigma_{{\gamma}d\to K^-K^+np}(E_{\gamma})}
{d{\bf {\Omega}}_{K^-}dp_{s}d{\bf {\Omega}}_s},
\end{equation}
%formula(44)
\begin{equation}
\frac{d\sigma_{{\gamma}d\to K^-K^+np}(E_{\gamma})}
{d{\bf {\Omega}}_{K^-}}=
\int\int{dp_{s}}{d{\bf {\Omega}}_{s}}
\frac{d\sigma_{{\gamma}d\to K^-K^+np}(E_{\gamma})}
{d{\bf {\Omega}}_{K^-}dp_{s}d{\bf {\Omega}}_{s}}.
\end{equation}
   
      Let us discuss now the results of our calculations in the framework
of the approach outlined above.
 
\section*{3 Results}

   At first, we consider the exclusive differential cross section (38) for
the process $d(\gamma, K^-p)K^+n$ proceeding through the intermediate
$\Theta^+$ state. There are many options to display the information contained
in this cross section. In particular, in fig. 4 we show the exclusive 
$K^-$-meson differential cross sections in the lab frame for the
proton--spectator emerging in the direction of the incoming photon
(i.e. at ${\bf {\Omega}_s}={\bf {\Omega}_{\gamma}}$, where
${\bf {\Omega}_{\gamma}}={\bf p_{\gamma}}/p_{\gamma})$ with momentum of
$270~{\rm {MeV/c}}$ calculated by eq. (38) for different assumptions concerning
the parity of the $\Theta^+$ state and the availability of the limitations
(32), (36) we introduced on the phase space of the $K^-p$ system at beam energy
of 1.5 GeV. The same as in fig. 4 but calculated for the photon energy of
1.75 GeV and the proton--spectator momentum of $45~{\rm {MeV/c}}$ is shown in
fig. 5. A choice of these two options for the incident energy and the
three--momentum of the proton--spectator has been particularly motivated by
the fact that the excess energies above the $K^-\Theta^+$ threshold, 
corresponding to both options (26 and 42 MeV, respectively), fall in the
near--threshold energy region of interest ($\le 100~MeV$). Under this choice,
the respective antikaon velocities in the ${\gamma}n$ c.m.s. turn out to be
smaller than the ones of this system in the lab frame. This means that in the
chosen kinematical conditions only the second term in eq. (38) plays a role
and, therefore, the $K^-$-meson production angle $\theta_{K^-}$ in the lab
system must be limited
\footnote
{Since in the chosen kinematics $\theta_{K^-}=\theta_{K^-}^c$ and the angle
$\theta_{K^-}^c$, defined above by eq. (40), is limited in line with the text
given just below the eq. (42).}
.   
Our calculations show that the maximal value of this angle amounts to $28.7^0$
and $26.3^0$ for $E_{\gamma}=1.5$ GeV, $p_s=270~{\rm {MeV/c}}$,
${\bf {\Omega}_s}={\bf {\Omega}_{\gamma}}$ and $E_{\gamma}=1.75$ GeV, 
$p_s=45~{\rm {MeV/c}}$, ${\bf {\Omega}_s}={\bf {\Omega}_{\gamma}}$,
respectively, which is reflected in the results we have exhibited in figs. 4
and 5. Looking at these figures, one can see that there are a clear differences
between the antikaon angular distributions calculated for different $\Theta^+$
parities and the same assumptions concerning the availability of the
limitations on the phase space of the $K^-p$ system (between dashed and solid,
double--dot--dashed and dot--dashed lines). Namely, in the case of 
negative--parity $\Theta^+$, the distributions (dashed and double--dot--dashed
curves) are strongly suppressed at forward angles $\theta_{K^-} \le 15^0$,
whereas in the case of positive--parity $\Theta^+$, the ones (solid and
dot--dashed lines) are flat at these angles. Moreover, although at larger
angles the respective differential cross sections belonging to the calculations
for different $\Theta^+$ parities have a similar shapes (compare dashed and
solid, double--dot--dashed and dot--dashed lines in figs. 4 and 5), their
strengths here for the negative--parity $\Theta^+$ are about forty times
smaller than those for the positive--parity one. Comparing the curves,
corresponding to the calculations for the same $\Theta^+$ parities with and
without placing the cuts under consideration on the phase space of the $K^-p$
system (solid and dot--dashed, dashed and double--dot--dashed lines,
respectively), one can also see that these cuts only slightly reduce the
strengths of the cross sections practically at all allowed angles in the case
of the kinematics of fig. 4, while in the kinematical conditions of fig. 5 they
decrease the cross sections only in small region of angles near the maximal
$K^-$-meson production angle
\footnote
{It is interesting to note that placing the cut (24) on the phase space of the
$K^+K^-$ system results in reduction of the antikaon yield from the process (1)
by the factors of about 1.7 and 1.5 in the kinematical conditions,
respectively, of fig. 4 and fig. 5. This means that about 40\% and 30\% of the
total $K^+K^-$ phase space are eliminated due to the cut (24), correspondingly,
in the former and the latter cases and, therefore, the larger part of this
space is free from the $\phi$-meson background.}
. 
This means that the main strengths
of the exclusive $K^-$-meson differential cross sections under consideration
concentrate in those parts of the sampled phase space where contribution
from the background sources, associated both with the $\Lambda(1520)$
production and the $K^-p$--FSI effects, is expected to be negligible in the
chosen kinematics. So, the foregoing shows that the observation of the
exclusive antikaon angular distributions from the process $d(\gamma, K^-p)K^+n$
proceeding via the intermediate $\Theta^+$ state near the threshold, like those
just considered, can serve as an important tool to distinguish the parity of
the $\Theta^+$ baryon.                                               

    Let us concentrate now on the exclusive differential cross section (43)
for the process $d(\gamma, K^-p)K^+n$ going through the virtual $\Theta^+$
state. In fig. 6 we show the exclusive $K^-$-meson differential cross sections
in the lab frame  for the proton--spectator emerging in the direction of the
incoming photon calculated by eq. (43) with allowance for the same scenarios
for the $\Theta^+$ parity and the availability of the limitations on the phase
space of the $K^-p$ system as in the preceding case at incident energy of
1.5 GeV. It can be seen that here also there are a distinct differences 
between the respective negative--parity $\Theta^+$ and the positive--parity
$\Theta^+$ results analogous to those observed previously. Namely, the
calculations including negative--parity $\Theta^+$ lie considerably lower the
positive--parity $\Theta^+$ results and, furthermore, their strengths are
substantially suppressed at forward angles $\theta_{K^-} \le 15^0$, while the
positive--parity $\Theta^+$ results are basically constant at these angles.
On the other hand, as shown in fig. 6, the differences between the calculations
for the same $\Theta^+$ parities with and without placing the cuts on the
phase space of the $K^-p$ system are small (cf. figs. 4 and 5), which means
that the main strengths of the exclusive antikaon differential cross sections
considered here lie in those parts of the sampled phase space where
contribution from the background sources, associated with the $\Lambda(1520)$
production and the $K^-p$-rescattering in the final state, is expected to be
negligible in the chosen kinematics. Therefore, the preceding gives the
opportunity to determine the $\Theta^+$ parity experimentally also by measuring
the exclusive antikaon angular distribution from the reaction 
${\gamma}d \to K^-\Theta^+p \to K^-K^+np$, like that just considered, in the
threshold energy region.

       Finally, let us focus on the inclusive differential cross section (44)
for the process $d(\gamma, K^-)K^+np$ proceeding through the intermediate
$\Theta^+$ state. In fig. 7 we show the inclusive $K^-$-meson differential
cross sections in the lab frame calculated by eq. (44) with employing the same
scenarios for the parity of the $\Theta^+$ pentaquark and the availability of
the limitations on the phase space of the $K^-p$ system as those of 
figs. 4, 5, 6 at the photon energy of 1.5 GeV. The same as in fig. 7 but
calculated for the photon energy of 1.75 GeV is shown in fig. 8. One can see 
that the distinctions between the corresponding negative--parity $\Theta^+$
and the positive--parity $\Theta^+$ calculations are quite clear both for 1.5
and 1.75 GeV initial energies and analogous to those observed before.
In particular, the cross sections calculated assuming that the $\Theta^+$ has
negative parity also are much suppressed at forward angles 
$\theta_{K^-} \le 15^0$, while the ones obtained supposing that the $\Theta^+$
has positive parity, as in the preceding cases, are practically constant at
these angles. Furthermore, although at larger angles
\footnote
{Restricted at $E_{\gamma}=1.5$ GeV as shown in fig. 7.}
the respective antikaon angular distributions belonging to the calculations for
different $\Theta^+$ parities also have a similar shapes (compare dashed and
solid, double--dot--dashed and dot--dashed curves in figs. 7 and 8), their
strengths here for the negative--parity $\Theta^+$ are significantly reduced
compared to those for the positive--parity one (cf. figs. 4, 5, 6). On the
other hand, the differences between the calculations for the same $\Theta^+$
parities with and without placing the cuts on the phase space of the $K^-p$
system, as in the preceding cases,  are largely insignificant, which means
that also the main strengths of the inclusive $K^-$-meson differential cross
sections under consideration concentrate in those parts of the sampled phase
space where contribution from the background sources, associated with
the $\Lambda(1520)$ production and the $K^-p$--FSI effects, is expected to be
negligible in the chosen kinematics. Therefore, the foregoing shows that the
inclusive antikaon angular distribution from the reaction
${\gamma}d \to K^-{\Theta^+}p \to K^-K^+np$ near the threshold also can be
useful to help determine the parity of the $\Theta^+$ pentaquark.

      Taking into account the above considerations, we come to the conclusion
that both the exclusive and inclusive $K^-$-meson laboratory differential
cross sections for the reaction 
${\gamma}d \to K^-{\Theta^+}p \to K^-K^+np$ near the threshold may be an
important tool to determine the parity of the $\Theta^+$ baryon. These
observables might be measured on modern experimental facilities such as
SPring--8, JLab, ELSA and ESRF.

\section*{4 Conclusions}

In this paper we have investigated in a spectator model the possibility
of determining the parity of the $\Theta^+$ pentaquark from the reaction
${\gamma}d \to K^-{\Theta}^+p \to K^-K^+np$ near the threshold. The elementary
$\Theta^+$ production process included in our study is
${\gamma}n \to K^-\Theta^+$. Taking into account the fact, established by the
authors of refs. [58, 59], that the free c.m.s. differential cross section for
this elementary process shows a clear distinction between the two opposite
parities of the $\Theta^+$ baryon close to the threshold and using their
predictions for this cross section, we have calculated the exclusive and
inclusive laboratory angular distributions of $K^-$-mesons produced through
this process, taking place on the moving neutron in the deuteron, for two
possible parity states of the $\Theta^+$ resonance at 1.5 and 1.75 GeV beam
energies with and without placing the relevant kinematical cuts on those parts
of the sampled phase space where contribution from the main background sources,
associated with the $\phi(1020)$, $\Lambda(1520)$ production as well as with
the $K^-p$--FSI effects, is expected to be dominant. We have shown that these
cuts play an insignificant role in the chosen kinematics, namely, they only
slightly reduce the antikaon angular distributions of interest, which means
that, in the chosen kinematical conditions, the main strengths of these
distributions concentrate in those parts of the sampled phase space where
the non--resonant background is expected to be negligible. On the other hand,
the calculated $K^-$-meson angular distributions were found to be strongly
sensitive to the $\Theta^+$ parity. We, therefore, come to the conclusion
that the observation of the exclusive and inclusive antikaon angular
distributions from the reaction ${\gamma}d \to K^-{\Theta}^+p \to K^-K^+np$
near the threshold can serve as an important tool to distinguish the parity
of the $\Theta^+$ pentaquark.
Such observation might be conducted at current experimental facilities.

  The author is grateful to A.I.Reshetin for interest in the work.

\newpage                                                                

%%%%%%%%%%%%%%%%%%%%%%%%%%%%%%%%%%%%%%%%%%%%%%%%%%%%%%%%%%%%%%%%%%%%%%%%%%%%%%%%
                                        
\begin{figure}[h!]
\centerline{\epsfig{file= 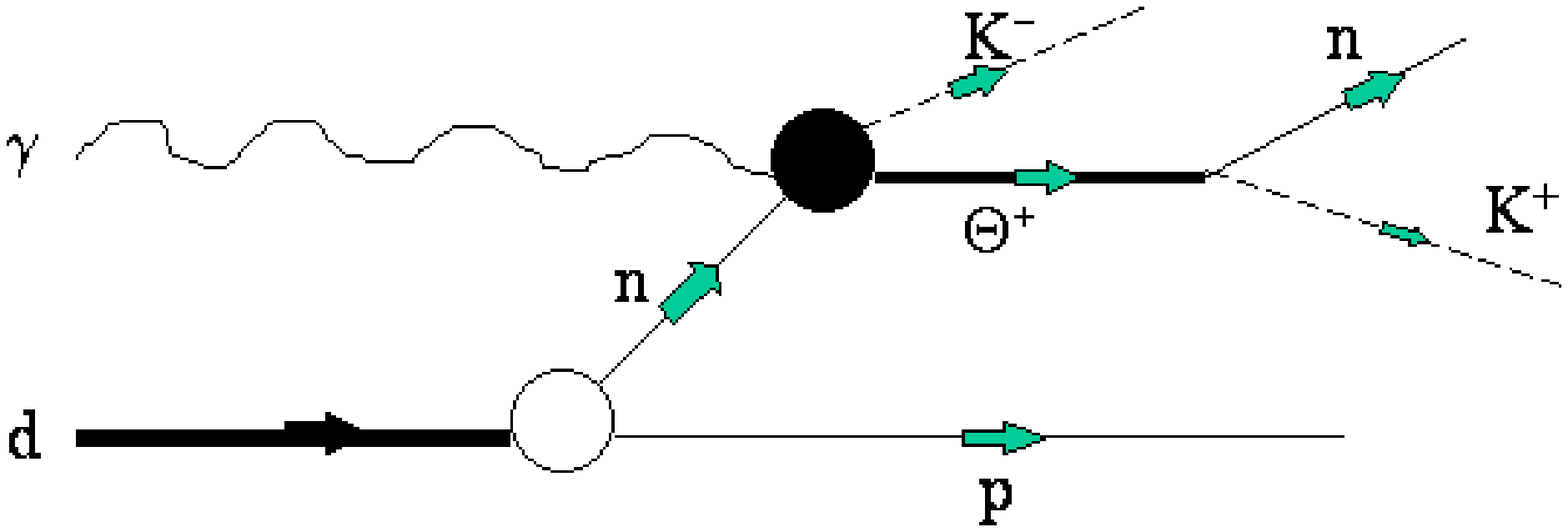,width=.66\textwidth,silent=,
clip=}}
\caption{\label{centered} Diagrammatic representation of $\Theta^+$ 
photoproduction from the deuteron within the Impulse Approximation 
  (the Spectator mechanism).}
\end{figure}

%%%%%%%%%%%%%%%%%%%%%%%%%%%%%%%%%%%%%%%%%%%%%%%%%%%%%%%%%%%%%%%%%%%%%%%%%%%%%

\begin{figure}[h!]
\centerline{\epsfig{file= 
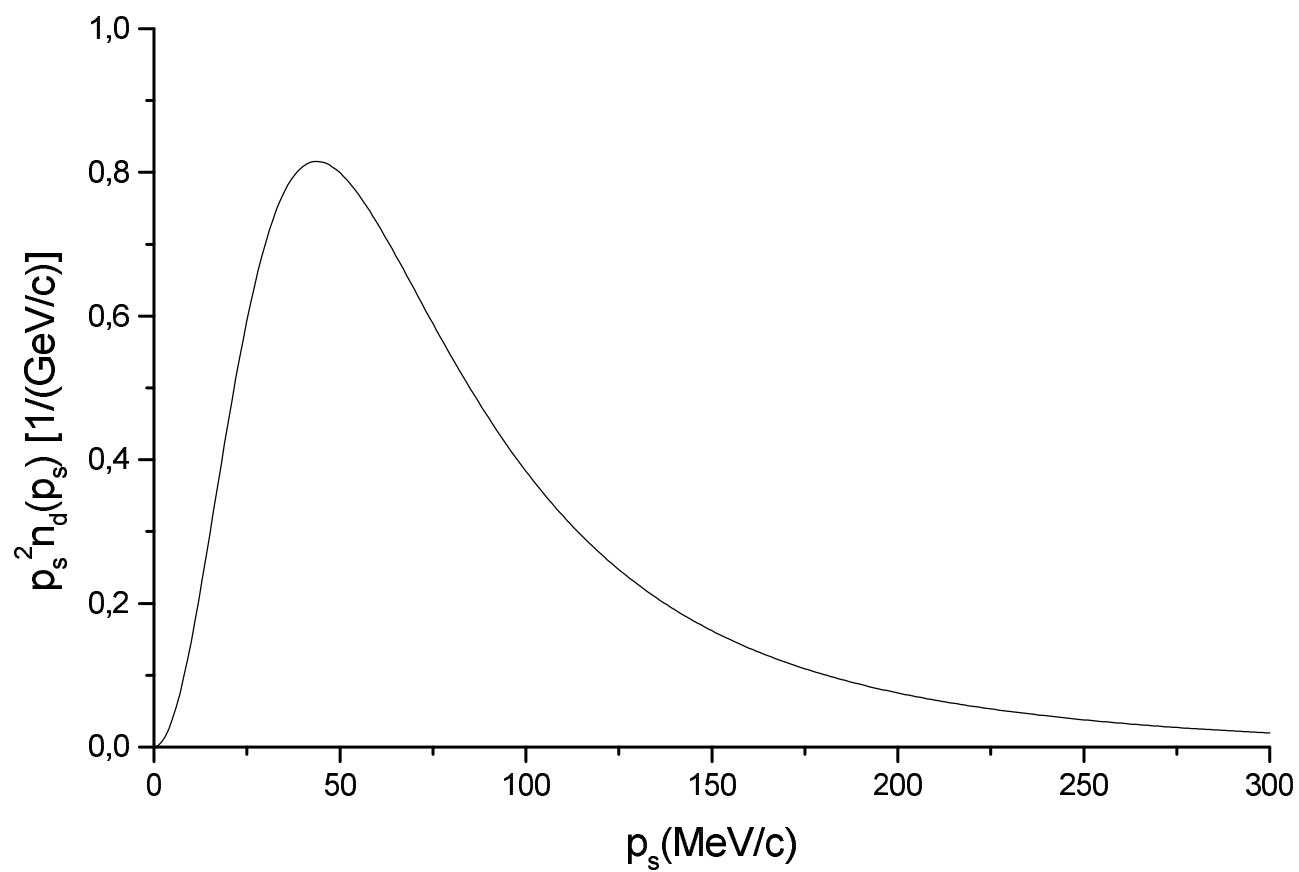,width=.88\textwidth,angle=270,silent=,
clip=}}
\caption{\label{centered}
Momentum distribution of proton--spectator in 
the reaction
  ${\gamma}d \to K^-{\Theta}^+p \to K^-K^+np$.}
\end{figure}

%%%%%%%%%%%%%%%%%%%%%%%%%%%%%%%%%%%%%%%%%%%%%%%%%%%%%%%%%%%%%%%%%%%%%%%%%%%%%%

\begin{figure}[h!]
\centerline{\epsfig{file= 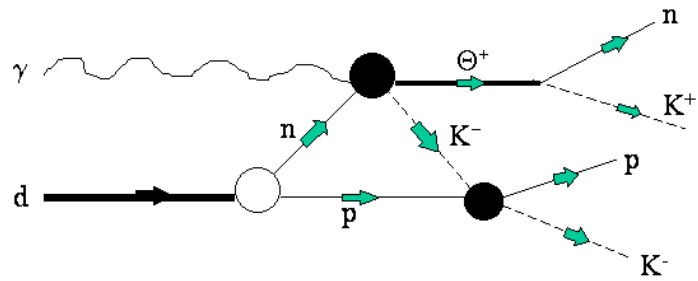,width=.66\textwidth,silent=,
clip=}}
\caption{\label{centered} Diagrammatic representation of $\Theta^+$ 
photoproduction
  from the deuteron including $K^-p$-rescattering in the final state.}
\end{figure}

%%%%%%%%%%%%%%%%%%%%%%%%%%%%%%%%%%%%%%%%%%%%%%%%%%%%%%%%%%%%%%%%%%%%%%%%%%%%%

\begin{figure}[h!]
\centerline{\epsfig{file= 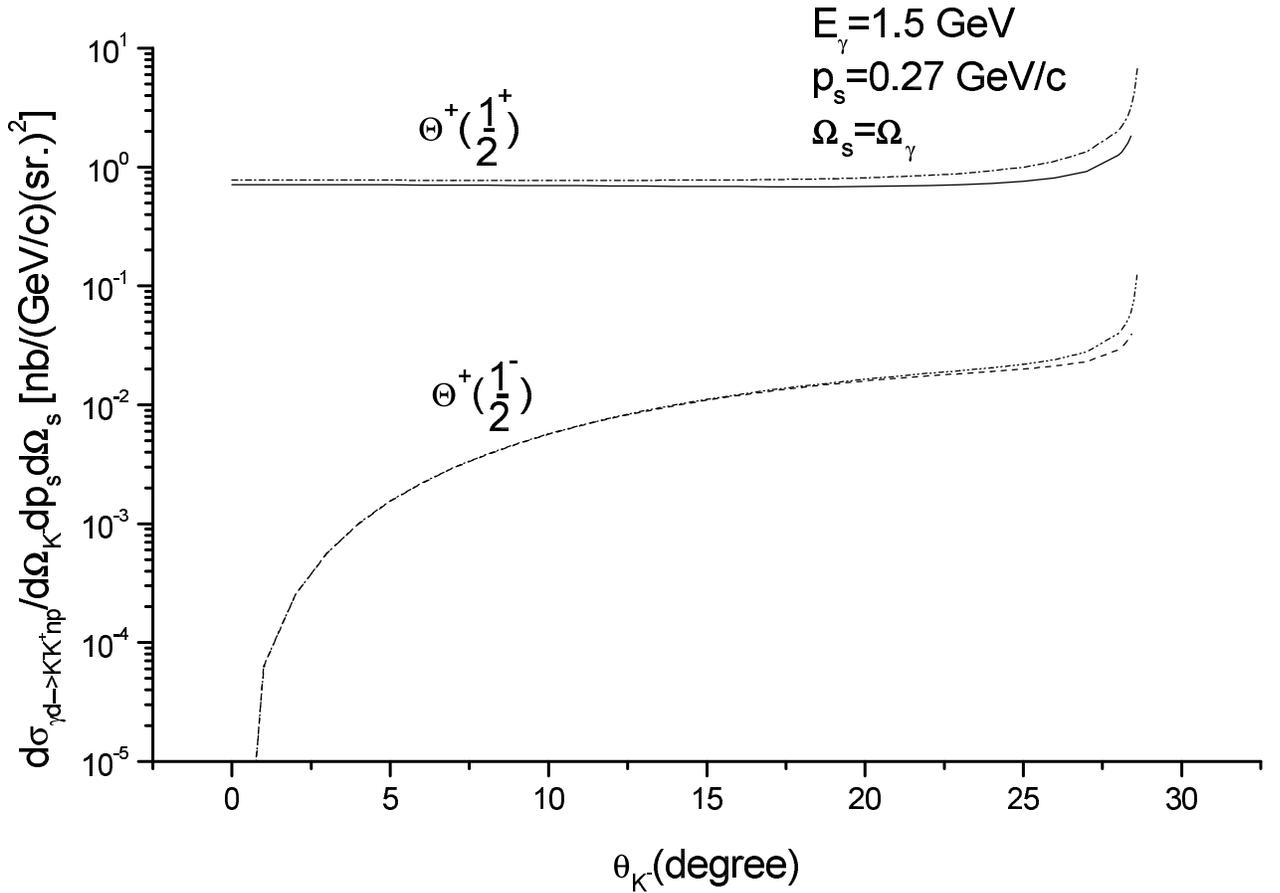,width=.88\textwidth,angle=270,
silent=,clip=}}
\caption{\label{centered} The exclusive $K^-$-meson differential cross 
  sections for 
  the process $d(\gamma, K^-p)K^+n$ proceeding through the intermediate
  $\Theta^+$ state at initial energy of 1.5 GeV with the proton--spectator
  emitted in the direction of an incident photon with momentum of 
  $270~{\rm MeV/c}$ as functions of the $K^-$ production angle in the
  nuclear lab frame. The solid and dot--dashed (dashed and double--dot--dashed)
  lines are calculations for positive (negative) $\Theta^+$ parity,
  respectively, with and without taking into account the limitations under
  consideration on the phase space of the $K^-p$ system. The limitation we
  introduced on the phase space of the $K^+K^-$ system is included in all
  calculations.}
\end{figure}

%%%%%%%%%%%%%%%%%%%%%%%%%%%%%%%%%%%%%%%%%%%%%%%%%%%%%%%%%%%%%%%%%%%%%%%%%%

\begin{figure}[h!]
\centerline{\epsfig{file= 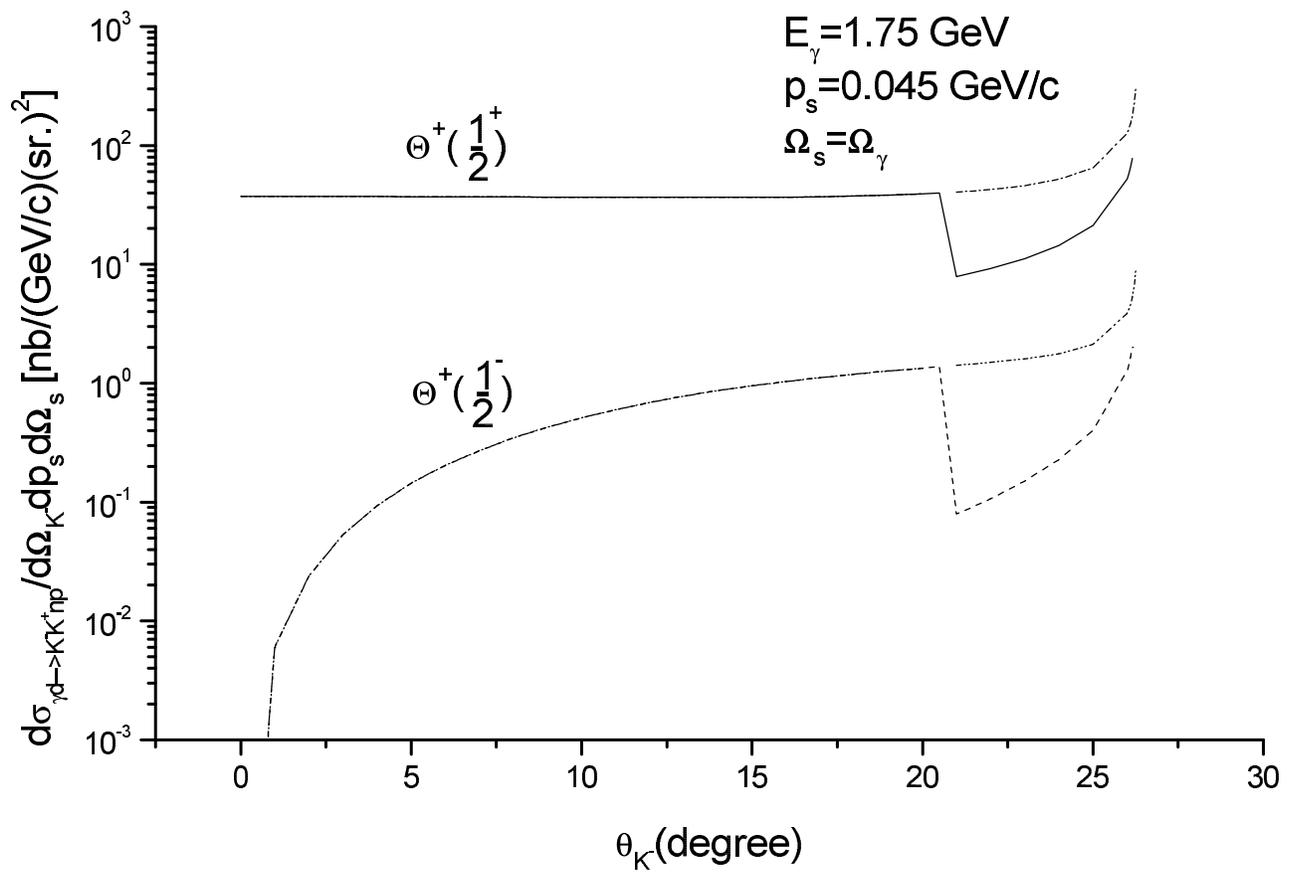,width=.88\textwidth,angle=270,
silent=,clip=}}
\caption{\label{centered} The same as in fig. 4 but for $1.75~{\rm GeV}$ 
beam energy and $45~{\rm MeV/c}$ proton--spectator momentum.}
\end{figure}

%%%%%%%%%%%%%%%%%%%%%%%%%%%%%%%%%%%%%%%%%%%%%%%%%%%%%%%%%%%%%%%%%%%%%%%%%%

\begin{figure}[h!]
\centerline{\epsfig{file= 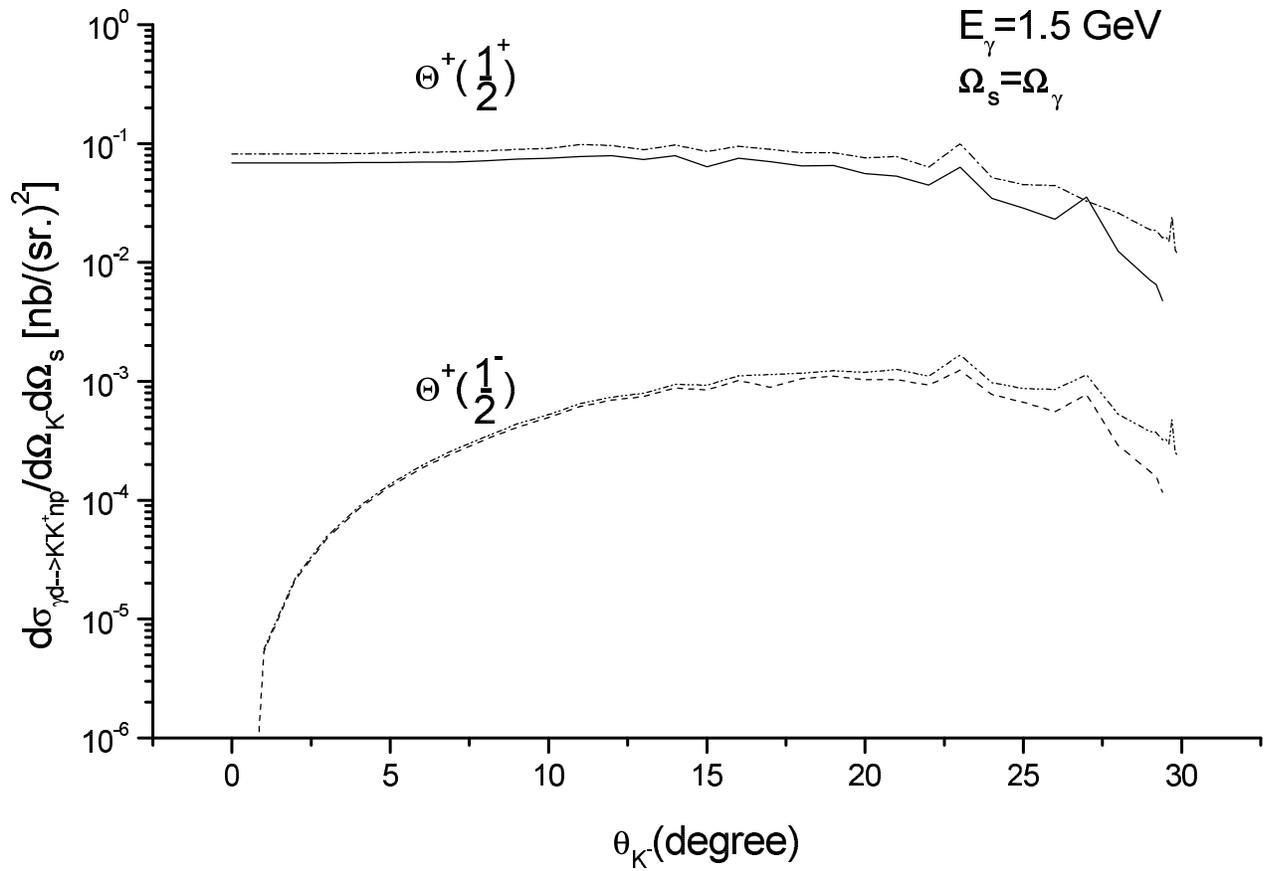,width=.88\textwidth,angle=270,
silent=,clip=}}
\caption{\label{centered} The exclusive $K^-$-meson differential cross 
  sections for
  the process $d(\gamma, K^-p)K^+n$ proceeding through the intermediate
  $\Theta^+$ state at initial energy of 1.5 GeV with the proton--spectator
  emitted in the direction of an incident photon 
  as functions of the $K^-$ production angle in the nuclear lab frame. The 
  notation of the curves is identical to that in fig. 4.}
\end{figure}

%%%%%%%%%%%%%%%%%%%%%%%%%%%%%%%%%%%%%%%%%%%%%%%%%%%%%%%%%%%%%%%%%%%%%%%%%%%

\begin{figure}[h!]
\centerline{\epsfig{file= 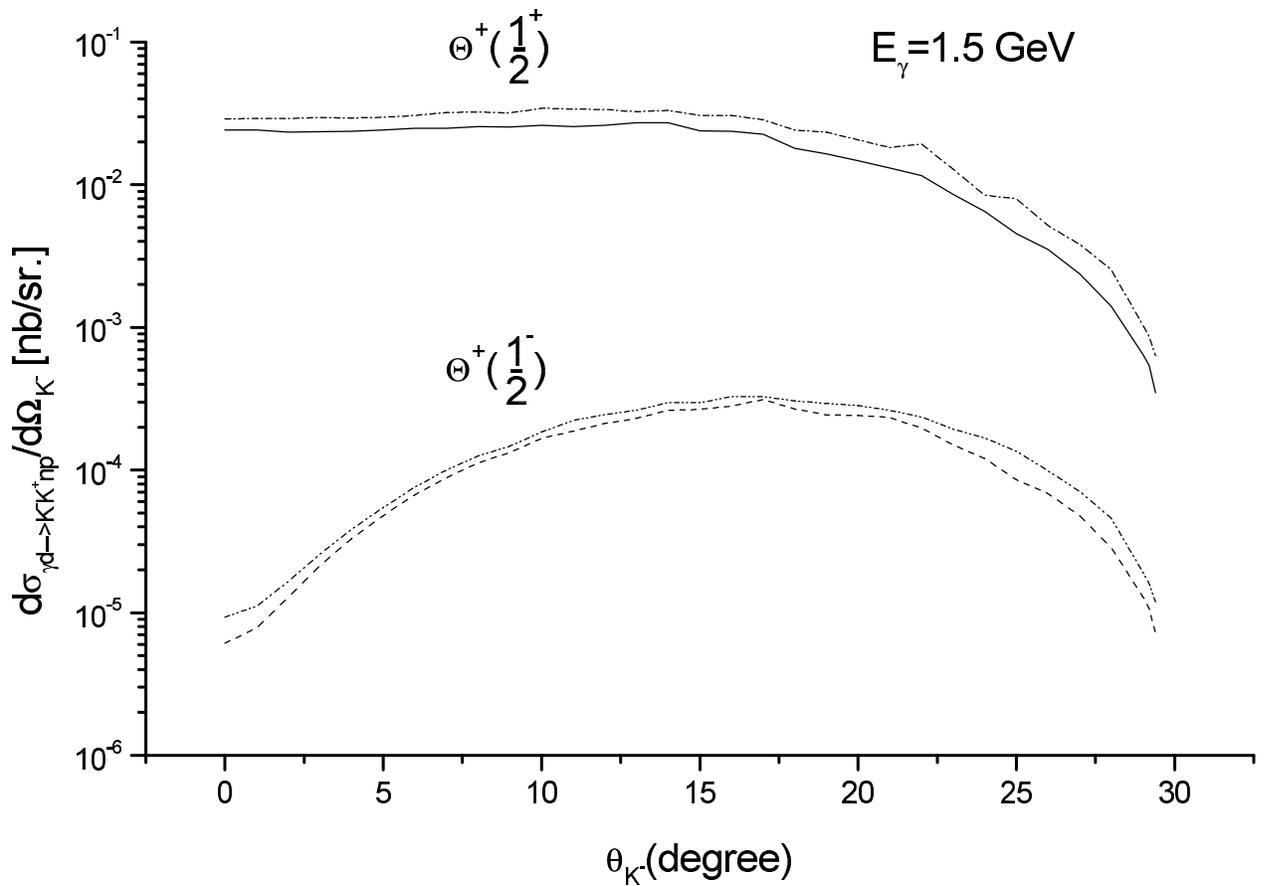,width=.88\textwidth,angle=270,
silent=,clip=}}
\caption{\label{centered} The inclusive $K^-$-meson differential cross 
  sections for
  the process $d(\gamma, K^-)K^+np$ proceeding through the intermediate
  $\Theta^+$ state at initial energy of 1.5 GeV 
  as functions of the $K^-$ production angle in the nuclear lab frame. The 
  notation of the curves is identical to that in fig. 4.}
\end{figure}

%%%%%%%%%%%%%%%%%%%%%%%%%%%%%%%%%%%%%%%%%%%%%%%%%%%%%%%%%%%%%%%%%%%%%%%%%%%

\begin{figure}[h!]
\centerline{\epsfig{file= 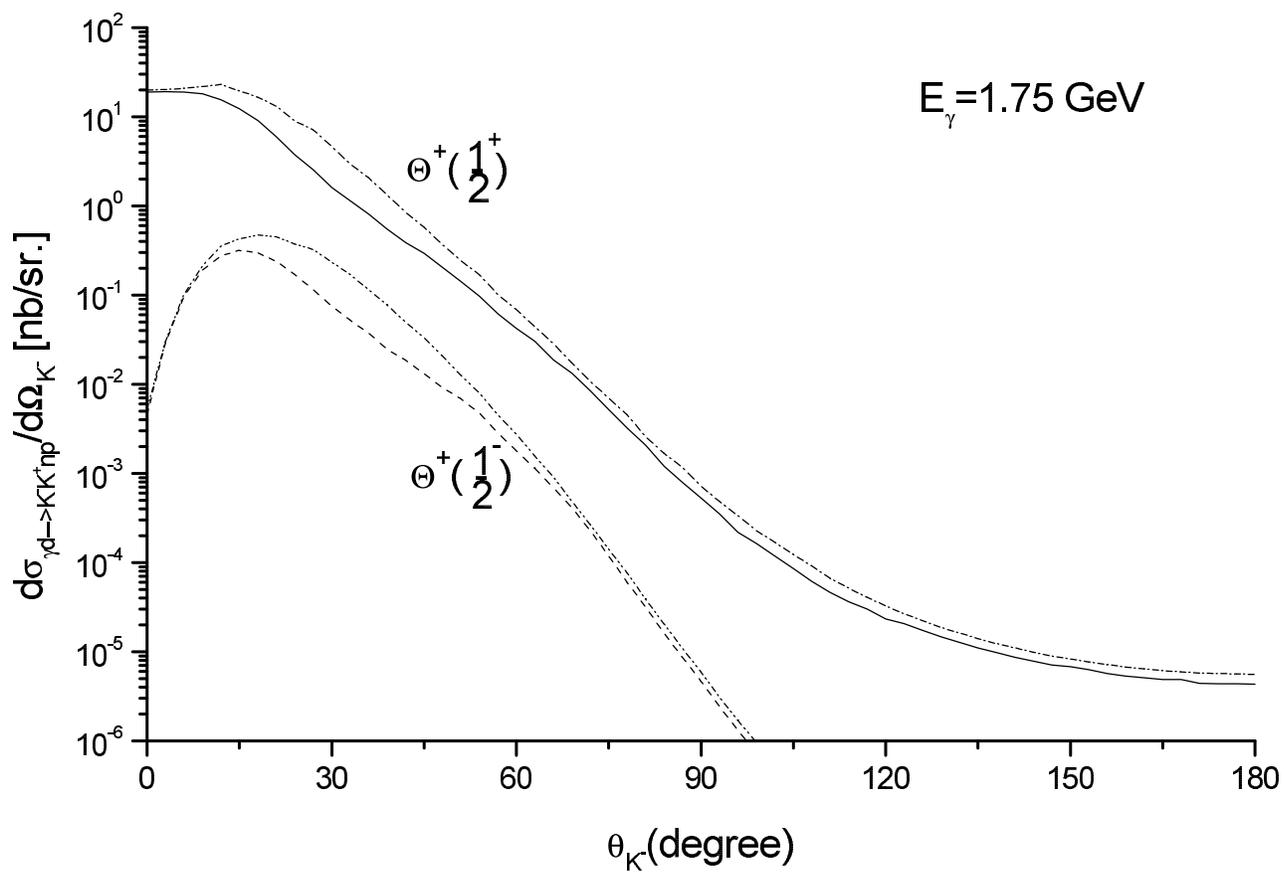,width=.88\textwidth,angle=270,
silent=,clip=}}
\caption{\label{centered} The same as in fig. 7 but for $1.75~{\rm GeV}$ 
beam energy.} 
\end{figure}

\end{document}